\documentstyle[aps,preprint]{revtex}
\begin{document}
\draft

\title{Anomalous Anharmonicity in Doped Cuprate Superconductors}
\author{V. Rzhevskii$^\star$ 
and G. Roepke$^{\star\star}$}
\address{$^\star$M.V.Lomonossov Moscow State University, Physics Department, 
119899 Moscow, Russia}
\address{$^{\star\star}$University of Rostock, FB Physik, 
Universitaetsplatz 1, 18051 Rostock, Germany}
\date{\today}
\maketitle

\begin{abstract}
Strong nonlinear elastic effects arise due to the Lifshitz transition
associated with a change in the Fermi surface topology. It is shown that an
electronic contribution to the elastic characteristics of cuprates becomes
important if the chemical potential is near but not too near the Van Hove
critical energy $\varepsilon_c$. In the case of a saddle point
singularity in the energy spectrum an enhancement factor arises for the ''electronic'' anharmonic constants of the 3rd and 4th order, that is proportional 
to $\varepsilon_F/(\varepsilon_F-\varepsilon_c)$ to the 1st and 2nd power, 
respectively. The behaviour of the Grueneisen coefficient in the vicinity 
of the Lifshitz transition is considered.
\end{abstract}
\vspace*{.5cm}

\pacs{PACS 74.25.Ld Mechanical and acoustical properties, elasticity, and ultrasonic attention, 74.25.Jb Electronic structure, 74.72.Bk Y-based cuprates, 74.72.Dn La-based cuprates.\\
Keywords: Electronic topological transition; Oxygen content; Grueneisen parameter; High temperature superconductors.}

High $T_{c}$ cuprate superconductors show unusual elastic properties in
the normal as well as in the superconducting states. There is experimental
evidence of a lattice softening in the normal state of $La_{2-x}Sr_{x}
CuO_{4}$ $(LSCO)$ at $x=0.14$ \cite{Nohara}. At the same time an unexpected 
lattice stiffening has been observed in the superconducting state. For
$YBa_{2}Cu_{3}O_{x}\, (YBCO)$ the values of the Grueneisen
parameter determined by ultrasonic techniques are spread over an unusually
broad range \cite{Can}. The decrease of the Grueneisen parameter with $x$ observed 
in \cite{Can} was recently confirmed by the oxygen concentration 
dependence of Raman active phonons \cite{Hir}. It is of interest to consider 
the unusual behaviour of the Grueneisen coefficient as a function of the 
oxygen content. As we will show below, a reason for such dependence may be 
a change of electronic structure with doping. In this work we demonstrate how the elastic properties depend on the peculiarities of the electronic structure of the doped cuprate superconductors. In particular we investigate 
the modification of the thermodynamic and elastic properties due to 
the change of the Fermi surface (FS) topology in the nonsuperconducting 
metallic state.\\ 
\indent An essential aspect of the electronic band structure of the cuprate superconductors is the quasi two-dimensionality. Van Hove  \cite{vHs} has shown that, in the 2D case,
inside each band the electron density of states has at least one infinite
logarithmic singularity (at the critical energy $\varepsilon _{c}$). The
influence of the Van Hove singularities on the properties of the high-$T_{c}$
superconductors has been widely studied \cite{etc.}. The effect of 
a Lifshitz electronic topological transition (LETT) in the 2D electron system
of cuprates has also attracted considerable attention \cite{Freem}, \cite{Onuf},
as this presents the possibility for a unified approach to studying the anomalies
in both normal and superconducting states. The LETT occurs when the Fermi energy
$\varepsilon _{F}$ passes through $\varepsilon _{c}$, and a change in the FS 
topology takes place \cite{Lif1}. The LETT in the 2D case for conventional metals
has been investigated in \cite{Ned}. The angular resolved photo-emission
experiments \cite{Pau} have revealed that $\varepsilon _{c}$ is located close
to $\varepsilon_{F}$ in the ''optimally'' doped cuprate superconductors. The
possibility of the Lifshitz transition in the cuprates was suggested by
Freeman \cite{Freem}, and it was considered recently in the high-$T_c$ superconductors by means of a microscopic $t-t'-J$ model, \cite{Onuf}.\\
\indent The metallic properties of cuprates are highly peculiar leading up 
to a deviation from the Fermi liquid (FL) behaviour of electrons \cite{Ander}. In particular, a dependence of magnetic susceptibility and Hall's constant 
on temperature, as well as the linear dependence of the resistivity on $T$
for HTCS materials with  the hole conductivity have been observed. Different aspects of these anomalies have been studied in connection with the characteristic features of the high $T_{c}$ superconductors in normal phase, as it is customary to assume they define a behaviour of cuprates in superconducting state. 
There exists a pseudogap in the underdoped regime above $T_{c}$, a rise of the localized magnetic moments and strong magnetic fluctuations over a wide doping range outside an area of antiferromagnetic phase. These effects can be considered as a manifestation of the strong electron correlations in cuprates. Therefore, a simple mean field treatment is not possible in this case.\\
\indent At the same time the appearance of various anomalies in the thermal and electronic properties of the free electron system due to the Lifshitz transition may be responsible for the anomalies in the interacting electron system. In this work we wish to focus our attention on the anomalies of the thermodynamic and elastic characteristics of cuprates which come from the peculiarities of the 2D electronic structure and layered anisotropy. So, we suppose that the FL approach is a suitable starting point as applied to normal metallic state of cuprates at some distance from the ''optimal'' doping. Below we discuss the influence of the electron interaction and a renormalization of the electron spectrum in cuprates.\\
\indent In the conventional metals a topological transition 
leads to strong nonlinear effects \cite{Lif2}. One can expect that elastic
anomalies due to the LETT arise in the cuprate materials also. For simplicity,  our investigation is performed for the 2D case. As well, we restrict our consideration to the case of a simple saddle point in band structure of the YBCO and LSCO compounds.\\
\indent To derive the behaviour of the elastic constants in the metallic
normal-state of doped cuprates it is necessary to determine how a basic
parameter of the LETT, the difference $z=\mu-\varepsilon _{c}$,
depends on the strain tensor $u_{ij}$ ($i,j=1,2$) of the 2D crystal.
The renormalization of the chemical potential  $\mu$ (at $T=0$)
due to deformation has been given in \cite{Lif2} and it can be easily
generalized to the 2D case. Then, we obtain for the deformed crystal
\begin{equation}
z=z_{0}+\Lambda _{ij}u_{ij},\label{au}
\end{equation}
where $z_{0}=(\varepsilon_{F}-\varepsilon _{c})|_{{u_{ik}}=0}$,
$\Lambda_{ij}=\bar{\lambda}_{ij}-\lambda_{ij}$,
$\lambda_{ij}(\vec{k}) = \lambda_{ji}(\vec{k})$ is the tensor of the
deformation potential at $u_{ij}=0$, ($\vec{k}$ is an electron wave vector),
$\bar{\lambda}_{ij}$ denotes its average over the FS. The Lifshitz
transition point is defined by the condition $z=0$. We suppose that the term
$z_{0}$ can be varied by doping so that it becomes sufficiently small. In
addition, we consider an elastic field of deformation in the form of a sound
wave. Then the transition parameter $z$ in cuprates consists of
a static and a dynamic part \cite{Rzh}.\\
\indent The electron density of states (DOS)in the 2D system is given by the
expression
\begin{equation}
\nu (\varepsilon )=2S\frac{d}{d\varepsilon }\int \theta (\varepsilon
(\vec{p}) - \varepsilon)\frac{d^{2}\vec{p}}{(2\pi \hbar )^{2}},\label{bu}
\end{equation}
where $S$ is the area occupied by the electrons, $\theta(x)=1$ for $x<0$,
and $0$ else is a step-function. In the case of a simple saddle point, near
the Van Hove critical energy $\varepsilon_c$, in the 2D band, the electron
dispersion relation may be written as
\begin{equation}
\varepsilon (\vec{p})=\varepsilon _{c}+\frac{p_{x}^{2}}{2m_{1}^{\ast}}-
\frac{p_{y}^{2}}{2m_{2}^{\ast}},\label{cu}
\end{equation}
where $p_{x}, p_{y}$ are the projections of $\vec{p}$ and
$m_{1}^{\ast}, m_{2}^{\ast} >0$ are the effective masses. Substituting
Eq.\ (\ref{cu}) in Eq.\ (\ref{bu}), it may be easily seen that a logarithmic
singularity arises in $\nu(\varepsilon)$. We decompose the electron density
of states into a regular part $\nu_{0}(\varepsilon )$ and a singular part
$\delta\nu (\varepsilon )$:
$\nu (\varepsilon )=\nu _{0}(\varepsilon ) + \delta \nu(\varepsilon)$. The
singular part results from the integration over the Fermi surface in the
break-off region
\begin{equation}
\delta \nu (\varepsilon ) = -\beta \ln (|\varepsilon -\varepsilon
_{c}|/\varepsilon _{c}), \label{du}
\end{equation}
where $\beta=S\pi ^{-2}\hbar ^{-2}(|m_{1}^{\ast}m_{2}^{\ast}|)^{1/2}$. For
the sake of definiteness, we will consider the case when an equienergy
surface neck appears with increasing energy, then the sign of a singular part
of the electron density of states is positive, $\delta\nu (\varepsilon )>0$.
It is easy to verify that in the case of the neck disruption, as the energy
grows, a singular part $\delta \nu (\varepsilon)$ has the opposite sign
relatively to Eq.\ (\ref{du}), $\delta \nu (\varepsilon )< 0$.\\
\indent The free energy $F$ can be also decomposed into a regular and a
singular part: $F=F_0 + \delta F$. $\delta F$ due to electronic transition
may be represented in terms of $z=\mu - \varepsilon _{c}$, where $\mu$ is
the chemical potential, \cite{Ned}:
\begin{equation}
\delta F=\beta \frac{z^{2}}{2}\ln \frac{|z|}{\varepsilon _{c}}
\left[ 1+\frac{\pi ^{2}}{3}\frac{T^{2}}{z^{2}}\right], \label{eu}
\end{equation}
here we suppose that $\mu\gg z\gg T$ (temperature is in energy units).
$F_{0}$ is assumed to be an analytical
function of $z$. We may neglect the temperature dependence
of the elastic constants in the considered temperature range.\\
\indent The standard equations of the theory of elasticity are
\begin{equation}
\rho \partial ^{2}u_{i}/\partial t^{2}=\partial \sigma _{ij}/\partial x_{j},
\qquad\sigma _{ij}=\sigma _{ij}^{latt}+\sigma _{ij}^{\prime},
\end{equation}
where $\rho$ is a mass density, $u_i$ is a displacement
vector. The lattice term $\sigma _{ij}^{latt}$ holds the linear
contribution $\lambda _{ijlm}u_{ij}$, where $\lambda _{ijlm}$ is a tensor of
elastic moduli, and the conventional anharmonic terms of expansion. The
electronic term $\sigma _{ij}^{\prime}\sim\partial\delta F/\partial u_{ij}$
contains a logarithmic singularity,
\begin{equation}
\sigma _{ij}^{\prime }=\beta _{0}z\Lambda _{ij}\ln \frac{|z|}{
\varepsilon_{c}}+\frac{\beta _{0}}{2}z\Lambda _{ij}. \label{fu}
\end{equation}
The constant $\beta _{0}=\beta /V$ is of the order
$\sim \nu _{0} (\varepsilon _{F})/\pi ^{2}v_{0}$, where
$\nu _{0}\left( \varepsilon _{F}\right)$ is the regular part of density
on the FS, $v_{0}$ is the volume of the unit cell of the crystal.\\
\indent There are two different limiting cases from Eq.\ (\ref{au}): (i)
$|z_{0}|=|\varepsilon_{F} -\varepsilon_{c}|_{{u_{ij}}=0}$ is comparable with
$|\Lambda _{ij}u_{ij}|$ in magnitude, or (ii) far exceeds it. We consider
the latter case, when the sound wave propagates along the metallic phase of
the cuprates, and the values $\stackrel{}{u_{ij}}$ are such that
\begin{equation}
|z_{0}|\gg |\Lambda _{ij}u_{ij}|. \label{gu}
\end{equation}
Then, using the expansion of $\delta F$, Eq.\ (\ref{eu}), with respect
to the small parameter $|\Lambda _{ij}u_{ij}|/|z_{0}|\ll 1$, (here
the singular terms at $|z_{0}|/\varepsilon_{c}\ll 1$ are retained)
\begin{equation}
\delta F\cong \frac{\beta }{2}(\Lambda _{ij}u_{ij})^{2}\ln \frac{|z_{0}|}{
\varepsilon _{c}}\pm \frac{\beta }{6}
\frac{(\Lambda _{pq}u_{pq})^{3}}{|z_{0}|}-
\frac{\beta }{24}\frac{(\Lambda _{lm}u_{lm})^{4}}{z_{0}^{2}}, \label{hu}
\end{equation}
the electronic stress tensor, Eq.\ (\ref{fu}), may be expanded up to the
second order terms as:
\begin{equation}
\sigma _{ij}^{^{\prime }}=\beta _{0}\Lambda _{ij}\Lambda _{lm}u_{lm}
\ln \frac{|z_{0}|}{\varepsilon _{c}}\pm\frac{\beta _{0}}{2}\frac{
\Lambda _{ij}}{|z_{0}|}(\Lambda _{lm}u_{lm})^{2}. \label{iu}
\end{equation}
Here the upper sign corresponds to the region $z_0 > 0$ where the Fermi
surface neck exists, and a lower sign refers to region $z_0<0$ where the neck
has been disrupted. The quadratic term in Eq.\ (\ref{iu}) has the same form
as the lattice contribution due to the cubic anharmonicity. Using
Eq.\ (\ref{hu}) and the estimate $|\Lambda _{ij}|\sim\Lambda \delta_{ij}$,
where $\Lambda >0$ is of the order of the deformation potential constant, the
''electronic'' anharmonic constants of 3rd and 4th order, $\Gamma_{3}$,
$\Gamma _{4}$, can be written as
\begin{equation}
\Gamma _{3}\cong \pm\frac{\beta _{0}}{6}\Lambda ^{2}|\eta|,
\qquad \Gamma_{4}\cong -\frac{\beta _{0}}{24}\Lambda ^{2}\eta ^{2}, \label{ju}
\end{equation}
where $|\eta| =\Lambda /\left| z_{0}\right| \sim \varepsilon _{F}/\left|
\varepsilon _{F}-\varepsilon _{c}\right| \gg 1$; $\Lambda > \varepsilon_F$ as
we shall discuss below. The case considered here could be formally interpreted
as an increase of the anharmonic constants $\Gamma_{3},\Gamma _{4}$ with an
enhancement factor $\eta$, to the corresponding power. Thus the electronic
contribution to the anharmonicity increases as the LETT point is approached.\\
\indent In conventional metals $\Lambda$ is usually of the same order as
$\varepsilon_{F}$ \cite{Lif2}. Since a value of $\varepsilon_{F}$ in cuprates
is well below, at first sight it would seem that the influence of electrons
on elastic properties of these materials is modest. But in the metallic state
of cuprates, owing to a relatively low density of carriers and a poor
screening (compared to the standard metals), the Coulomb energy of interaction
is large, and can cause the value of $\Lambda$ to increase. The quasi-two
dimensionality of the electronic structure in cuprate superconductors also
provides an explanation for  $\Lambda$ increasing. For the quantitative
estimates of $\Lambda$ we used a familiar procedure from the method of
deformation potential \cite{Ziman}. Hence we take into account that the
electrons of conductivity in cuprates are basically localized in the
$CuO_{2}$ planes. Using an effective density of electronic states per atom
of $Cu$, \, $\nu_{0}\left(\varepsilon_{F}\right)\cong 4.8$
$(\rm {eV\cdot at.}Cu)^{-1}$ in the case of $YBCO$\, and \,
$\nu _{0}\left(\varepsilon_{F}\right)\cong 4.14$ $(\rm {eV\cdot at.}Cu)^{-1}$
for $LSCO$, \cite{Plakida}, we obtain $\Lambda \cong$1.25 eV and $\Lambda
\cong$ 0.82 eV, respectively, which are larger than the accepted values of
$\varepsilon_{F}$ in these systems.\\
\indent  The relation between the energy distance $z_{0}$ and the
doping parameter $x$ is of great importance \cite{Bok}. For simplicity,
we treat the dependence of $z_{0}$ on the doping $x$ as an empirical
fact. The dopant dependence of $z_{0}$ has been estimated from experimental
data for the chemical potential shift in \ doped $La_{2-x}Sr_{x}CuO_{4}$,
\cite{Ino}, and $YBa_{2}Cu_{3}O_{x}$, \cite{Liu}. From the above-mentioned
estimates we have obtained the following values of the ''electronic''
anharmonic constants in $LSCO$ for $|z_0|\sim 100$K:
$|\Gamma_{3}| \cong 1.33$GPa, \,$\Gamma _{4}\cong -30$GPa; in $YBCO$ for
$|z_0|\sim 250$K: $|\Gamma_{3}| \cong 14$GPa,\,$\Gamma _{4}\cong -189$GPa.
The dependence of ''electronic'' anharmonic elastic constants of 3rd and
4th order for compounds $LSCO$ and $YBCO$ on doping are shown in Fig.1 (b),
(d). A dependence of $\Gamma _{3}, \Gamma _{4}$ on doping, Eq.\ (\ref{ju}), and
the asymmetry of $\Gamma _{3}$ relatively to the point $z_{0}=0$ offer a reason for 
a distinction between a metallic phase of cuprates and the conventional metals in 
the kinetic characteristics and transport properties. Also, the anharmonic
dynamics is highly sensitive to the ion mass \cite{Plakida}, so that one
can expect a modification of isotope shift with doping.\\
\indent The Grueneisen constant is defined as the ratio of the full
coefficient of thermal expansion $\alpha _{V}$ to the full specific heat
of the crystal at constant volume $C_{V}$
\begin{equation}
\gamma =\frac{V\alpha _{V}B_{T}}{C_{V}}, \quad\mbox{or}\quad
\gamma =\frac{V}{T}\left( \frac{{\partial }^{2}F}{\partial T\partial V}
\right) /\frac{{\partial }^{2}F}{\partial T^{2}}, \label{ku}
\end{equation}
where $B_{T}$ is the isothermal bulk modulus and $F$ is the free energy. The
spectrum of oscillations in cuprates as anisotropic layered crystal structure
is characterized by more that one Debye temperature. In particular,
a temperature range, where the interaction between the layers is not
essential, is \cite{Lif3},
\begin{equation}
\left(E_{\bot }/E_{\Vert }\right) ^{1/2}\Theta \ll T\ll
\Theta ,\ \ \ \ E_{\bot }\ll E_{\Vert }, \label{lu}
\end{equation}
here $\Theta = {\pi\hbar s_{\Vert }}/a$ is a characteristic temperature,
corresponding to a velocity $s_{\Vert }$ of propagation of sound wave in the
$ab$-plane of layer $CuO_{2}$, $s_{\Vert }=(E_{\Vert }/\rho)^{1/2}$; $\ a$ is
a cell parameter, $E_{\Vert }$ and $E_{\bot }$ are the Young's moduli in the
plane $ab$ $\left( \lambda _{xxxx}\right) $ and along the $c$-axis $\left(
\lambda_{zzzz}\right) $ accordingly. Let us estimate the temperature range
defined by Eq.\ (\ref{lu}). In the case of $YBa_2Cu_3O_7$, using
$E_{\Vert} = 2.30\times 10^{11}\rm{Pa}$, $E_{\bot } = 1.60\times10^{11}
\rm{Pa}$ \cite{Saint-Paul}, and a mass density $\rho = 6.4 \rm{g/cm^3}$,
we obtain $\Theta = 356\rm{K}$ and $(E_{\bot}/E_{\Vert})^{1/2}\Theta = 296
\rm{K}$. For $LSCO$ respectively, using $E_{\Vert} = 2.48\times 10^{11}\rm
{Pa}$, $E_{\bot } = 2.05\times10^{11}\rm{Pa}$ \cite{Migliori}, $\rho =
6.88\rm{g/cm^3}$, we find $\Theta = 361\rm{K}$ and $(E_{\bot}/E_{
\Vert})^{1/2}\Theta = 328\rm{K}$. Notice that the temperature boundaries in
Eq.\ (\ref{lu}) in turn themselves depend on the doping.\newline
\indent A singular part of the lattice Grueneisen parameter $\delta\gamma^
{ph}$ can be obtained by estimating the corrections to the squared phonon
frequency averaged over the Brillouin zone, resulting from the topological
transition. Our estimates have shown that the latter are
relatively small. Much more sensitivity to the doping in the region of the LETT
was found for the electronic Grueneisen coefficient, $\gamma^{e}$. To
calculate $\gamma^{e}$, we shall present, as previously, the free energy as
a sum $F=F_0 + \delta F$, then $\gamma^{e}=\gamma^{e}_{0}+\delta\gamma^{e}$.\\
\indent The regular parts of the electronic specific heat $C_{V}$ and
the thermal coefficient of pressure are
\begin{equation}
C_{V}=\frac{\pi ^{2}}{3}\nu _{0}(\varepsilon _{F})T;\quad \quad
\left( \frac{\partial P}{\partial T}\right) _{V}=\frac{\pi ^{2}}{3}
\frac{\partial \nu _{0}(\varepsilon _{F})}{\partial V}T. \label{mu}
\end{equation}
According to Eq.\ (\ref{eu}), the singular part of the electronic specific
heat follows as
\begin{equation}
\delta C_{V}=-T \frac{\partial ^{2}\delta F}{\partial T^{2}}\cong - \frac{
\pi^{2}\beta}{3}T{\mbox ln}\frac{|\eta|\varepsilon _{c}}{\Lambda}. \label{nu}
\end{equation}
The singular part of the thermal coefficient of pressure is
\begin{equation}
\frac{1}{T}\delta\left( \frac{\partial P}{\partial T}\right) _{V} =
-\frac{1}{T}\frac{\partial ^{2}\delta F}{\partial T\partial V}\cong -
\frac{\pi ^{2}\beta_{0}}{3}\eta. \label{ou}
\end{equation}
Using Eqs.\ (\ref{ku}), \ (\ref{mu})--\ (\ref{ou}) and detaching a standard
electronic Grueneisen coefficient $\gamma _{0}^{e}=\frac{V}{\nu_{0}}\frac{
\partial \nu _{0}}{\partial V}=\partial \ln\nu _{0}/\partial \ln V$, we
obtain the electronic contribution to the Grueneisen parameter near the
point of the LETT as
\begin{equation}
\gamma ^{e}\cong \gamma _{0}^{e}/c(\eta)+\delta\gamma^{e}, \quad \delta
\gamma^{e}=-|\eta|/\pi^{2}c(\eta), \label{pu}
\end{equation}
where $c(\eta)=(1+\frac{1}{\pi ^{2}}\ln{\frac{|\eta|\varepsilon _{c}}
{\Lambda}})(1+ C^{ph}_V/C^{e}_V)$. The dependence of the singular 
electronic correction $\delta\gamma^{e}$ due to the proximity to the LETT 
on doping accordingly to Eq.\ (\ref{pu}) is plotted in Fig.1 (a), (c). Under
certain conditions its variations may lead to negative values of
the Grueneisen coefficient. In particular, important point is the relation
between the phonon and electronic specific heats $C_{V}^{ph}/C_{V}^{e}$
which enters in Eq.\ (\ref{pu}). At relatively high temperatures ($T>T_{c}$)
the phonon contribution to the specific heat $C^{ph}_{V}$ is, as a rule, 
greater than the electronic one, $C^{e}_{V}$. But in the temperature range
Eq.\ (\ref{lu}), where $C_{V}^{ph}\sim T/\Theta$ due to the layered
anisotropy \cite{Lif3}, $C^{ph}_{V}$ is comparable with $C^{e}_{V}$.
For the metallic normal state of $LSCO$ another low temperature interval
where $C^{e}_{V}\gg C^{ph}_{V}$ is available:
$T\ll \left(E_{\bot }/E_{\Vert }\right) ^{3/4}\Theta (\Theta/\varepsilon_F)
^{1/2}$. This case is the only one plotted in Fig.1 (a).\\
\indent Let us take up the assumptions underlying the results obtained
above. We assumed $\varepsilon _{F}\gg |z|\gg T$. The restriction
$\varepsilon _{F}\gg |z|$ is obvious from setting up a problem, since the
ratio $|z|/\varepsilon _{F}$ is a dimensionless parameter that defines the
distance to the transition point. From Eq.\ (\ref{gu}) it follows that
$|u_{ij}|\ll |z_{0}|/|\Lambda _{ij}|\sim |z_{0}|/\Lambda \ll 1$.
Thus, the present treatment is appropriate to ultrasound waves (with finite
amplitudes) as well as to long-wavelength phonons. In this connection of
special interest is a problem of interplay between the ''electronic''
anharmonicity in plane $ab$ due to the proximity to the Lifshitz transition
and a ''tilting'' mode in $LSCO$, or, in case of $YBCO$, a pyramidal apex
oxygen motion along the c-axis.\\
\indent For temperatures in considered $T \ll\Theta_{D}, z_{0}$ a boundary of
superconducting state gives a restriction from below, $T>T_{c}$. At $T\neq0$
the singularities caused by the LETT are smeared out. The width of the
temperature smearing is $\Delta z\sim T$. \,So, the obtained results are
valid at temperatures such that the smearing is small: $T/|z_{0}|\sim\Delta
z/|z_{0}|\ll 1$. The inequality $T/|z_{0}|\ll 1$ coincides with a condition
for a degeneracy of the electron gas in the normal metallic state of cuprates.
The Fermi--liquid approach in this case may be considered as a close
approximation if \quad $\lambda\ll (\Theta_{D}/T)^2$ \,\cite{Gins}, \quad
where $\lambda$ is an electron--phonon coupling constant. The simple
estimates of $\lambda$ for $YBCO$ and $LSCO$ show that
above mentioned requirement is fulfilled for these compounds at $\lambda
\simeq 1.3 \div 4.2$. While a behaviour of $\lambda$ depends on the details
of anharmonic interactions \cite{Allen}. Taking into account the
''electronic'' contribution to the anharmonic constants, Eq.\ (\ref{ju}),
one can find $\lambda$ as a function of temperature \cite{Hardy} and doping,
to refine the regions of the Fermi--liquid theory feasibility.\newline
\indent Up to now we considered a free electron system in the presence
of the simple saddle point in the electron spectrum. Now we discuss how the 
fact that the electrons in the cuprates are a strongly correlated system will influence the results. Recently in \cite{Onuf1} it was shown that in 2D 
interacting system the density of states deviates from the bare DOS. Particularly, for the doping range around the critical point corresponding to 
the LETT, it is characterised by a shifted logarithmic singularity and two 
jumps on the right in energy of the logarithmic divergence point. Ignoring 
these jumps \footnote{What means that we have restricted ourselves to the region of the overdoped regime.}, one can conclude that the obtained results are qualitatively not changed for a renormalized spectrum in the presence of interaction \footnote{A role of the jumps in the underdoped regime for elastic properties we consider elsewhere.}. The quantitative changes are appreciably defined by the magnitude of the deformation potential. At the same time the renormalization of the deformation potential in the system of strong correlated electrons remains to be solved and demands further investigation. In this casen, it is of special interest to extend the deformation potential theorem for the electron phonon coupling due to Bardin to the electron-spin fluctuation coupling as given by Schrieffer \cite{Schrieffer}.\\
\indent The comparison of the anharmonic constants of 3rd and 4th order, 
$\Gamma_{3},\Gamma _{4}$, Eq.(11), with
the experimental values is not possible because of lack of data. 
However there are indications that the results obtained with the use of
$\Gamma_{3},\Gamma _{4}$ are consistent with other experimental data. Fig.1 (c)
shows that a change of the singular correction, $\delta\gamma^{e}$, of
the electronic Grueneisen coefficient in $YBa_{2}Cu_{3}O_{x}$, 
as function of oxygen content $x$ (solid curve) is in qualitative agreement
with a variation of a mean acoustic Grueneisen parameter with $x$ taken from experimental data \cite{Can}. A closer comparison would require further
experimental data with monocrystal samples in the proximity to the ''optimal'' doping.  In the case of $La_{2-x}Sr_{x}CuO_{4}$, at $x$=0.14 the  experimentally
observed anomaly (negative value) of the in-plane $ab$ thermal expansion
coefficient $\alpha_{\Vert}$ \cite{Sarrao} gives one more evidence of
agreement between our results and experiment. When $\alpha_{\Vert}$ is expressed in terms of the Grueneisen parameter using Eqs.\ (\ref{ku}),
\ (\ref{pu}), we obtain $\alpha_{\Vert}<0$ in the required temperature
range.\\
\indent In summary, we have presented the dependence of the elastic
properties on the singularities of the 2D electronic energy spectrum in
the metallic normal state of the doped cuprate superconductors, as applied
to $LSCO$ and $YBCO$ compounds. It has been shown that the proximity to the point 
of a topological modification of the Fermi surface gives rise to 
the unusual ''electronic'' anharmonicity, which essentially depends on the doping and 
may exceed the usual lattice anharmonicity. The correction to the electronic Grueneisen coefficient 
due to this anomalous anharmonicity was found to be negative, what is in line with
the experimental data for $YBCO$ and $LSCO$. In superconducting state,
''electronic'' anharmonicity can be of importance in the context of an
enhancement of the electron-phonon coupling to explain high $T_{c}$
in cuprates.\\
\indent We would like to thank the Deutsche Forschungsgemeinschaft for
support of this work, Grant no.436-Rus-113/202/0/R,S. V.R. gratefully acknowledges the hospitality of the University of Rostock.

\newpage
\begin{figure}[hbt]
\vspace{.8cm}
\centerline{\hspace{-2.0cm}\parbox[]{6cm}{
\input gr1a.pic}}
\vspace{1.4cm}
\centerline{\parbox[]{6cm}{
\input gr2aa.pic}} 
\vspace{.5cm}
\label{F}
\end{figure}
\newpage
\begin{figure}
\caption[]{Dependence of singular correction $\delta\gamma^{e}$ on doping $x$ 
for $LSCO$ $(a)$ and for $YBCO$ $(c)$ near the Lifshitz transition point.
In $(b)$, $(d)$ the similar dependence of elastic constants $\Gamma_3$, 
$\Gamma_4$ is shown. Dashed lines separate the regions with $z_0>0$ (left),
$z_0<0$ (right). In $(c)$ the open circles are the experimental points for
a change of mean acoustic Grueneisen parameter with $x$ \cite{Can}. 
A dotted line is intended as a guide to the eye.}
\end{figure}

\begin{thebibliography}{99}
\bibitem{Nohara} M. Nohara, T. Suzuki, Y. Maeno {\it et al.}, Phys. Rev. B
{\bf 52}, 570, (1995).
\bibitem{Can} Q. Wang, G.A. Saunders, D.P. Almond {\it et al.}, Phys. Rev. B
{\bf 52}, 3711, (1995).
\bibitem{Hir} T. Hirata, Physica B { \bf 263-264}, 822,(1999). 
\bibitem{vHs} L. Van Hove, Phys. Rev. {\bf 89}, 1189, (1953). 
\bibitem{etc.} For a review see, e.g., R.S. Markiewicz, J. Phys.
Chem. Solids {\bf 58}, 1179, (1997).
\bibitem{Lif1} I.M. Lifshitz, Sov. Phys. JETP (USA) {\bf 11}, 1130, (1960). 
\bibitem{Ned} S.S. Nedorezov, Sov. Phys. JETP (USA) {\bf 24}, 1061, (1967). 
\bibitem{Pau} R. Liu, B.W. Veal, A.P. Paulikas {\it et al.}, Phys. Rev. B
{\bf 45}, 5614, (1992).
\bibitem{Freem} A.J. Freeman, J. Yu, C.L. Fu, Phys. Rev. B {\bf 36}, 7111,
(1987).
\bibitem{Onuf} F.Onufrieva, P.Pfeuty, and M.Kiselev Phys. Rev. Lett. {\bf 
82}, 2370, (1999); F.Onufrieva and P.Pfeuty, Phys. Rev. Lett. {\bf 
82}, 3136, (1999).
\bibitem{Ander} P.W.Anderson, Science {\bf 235}, 1196, (1987);
\bibitem{Lif2} I.M. Lifshitz, V.V. Rzhevskii, M.I. Tribelskii,  Sov. Phys.
JETP (USA) {\bf 54}, 810, (1981). 
\bibitem{Rzh} V.V. Rzhevskii, G. Roepke, in Proceedings of the International
Conference ''Problems of Condensed Matter Theory'',  Moscow, June 1997. 
\bibitem{Ziman} J.M. Ziman, {\it Electrons and Phonons} (Oxford, Clarendon
Press, 1963).
\bibitem{Plakida} N.M. Plakida, {\it High-Temperature Superconductors}
(Springer Verlag, 1995).
\bibitem {Bok} J. Bouvier and J. Bok Physica C {\bf 288}, 217, (1997); J.
Giraldo, R. Baquero, Physica C {\bf 257}, 160, (1996). 
\bibitem{Ino} A. Ino, T. Mizokawa, A. Fujimori {\it et al.}, Phys. Rev. Lett.
{\bf 79}, 2101, (1997).
\bibitem{Liu} R. Liu, B.W. Veal, C. Gu {\it et al.}, Phys. Rev. B {\bf 52},
553, (1995).
\bibitem{Lif3} I.M. Lifshitz, {\it Selected Works. Electronic Theory of
Metals} (Nauka, Moscow, 1994), 359. 
\bibitem{Saint-Paul} M. Saint-Paul, J.L. Tholence, H. Novel {\it et al.},
Solid State Commun. {\bf 69}, 1161, (1989); M. Saint-Paul and J.Y. Henry,
Solid State Commun. {\bf 72}, 685, (1989).
\bibitem{Migliori} A. Migliori, W.M. Visscher, S. Wong {\it et al.}, Phys.
Rev. Lett. {\bf 64}, 2458, (1990).
\bibitem{Gins} V.L. Ginsburg and E.G. Maximov, Supercond. Phys. Chem. Tech.
{\bf 5}, 1543, (1992).
\bibitem{Allen} P.B. Allen and B. Mitrovic, in {\it Solid State Physics},
edited by H. Ehrenreich, F. Seitz, and D. Turnbull (Academic, New York, 1982),
Vol. 37, p. 1.
\bibitem{Hardy} J.R. Hardy and J.W. Flocken, Phys. Rev. Lett. {\bf 60},
2191, (1988).
\bibitem{Onuf1} F.Onufrieva, P.Pfeuty, and M.Kisselev, J. Phys. Chem. Solids
{\bf 59}, 1853, (1998).  
\bibitem{Schrieffer} J.R.Schrieffer, J. of the Korean Phys. Soc. {\bf 31}, 1, (1997). 
\bibitem{Sarrao} J.L.Sarrao, D.Mandrus, and A.Migliori et al., Phys. Rev. B
{\bf 50}, 13125, (1994-II).
\end{thebibliography}
\end{document}